# Hyperfine-driven persistent currents in mesoscopic rings based on a 2D electron gas with Rashba spin-orbit interaction

Vitaliy A. Cherkaskiy[1,2], Sergey N. Shevchenko[1,2], Alexander S. Rozhavsky[2,3], and Israel D. Vagner[3]

[1] Physical-Technical Department, Kharkov State University,
4 Svobody Square, 310077 Kharkov, Ukraine

[2] B. Verkin Institute for Low Temperature Physics and Engineering, National Academy of Sciences of Ukraine, 47 Lenin Ave., 310164 Kharkov, Ukraine

[3] Grenoble High Magnetic Field Laboratory, Max-Planck-Institut für Festkörperforschung and Centre National de la Recherche Scientifique B. P. 166, 38042 Grenoble Cedex 09, France
E-mail: vagner@labs.polycnrs-gre.fr

We present a detailed theory of induced persistent current (PC) produced by hyperfine interaction in mesoscopic rings based on a 2D-electron (hole) gas in the absence of external magnetic field. PC emerges due to combined action of the hyperfine interaction of charge carriers with polarized nuclei, spin-orbit interaction and Berry phase.

### Introduction

The current situation in solid state physics is characterized in particular by a hectic search of various macroscopic topological quantum effects. The most popular of these are the persistent current (PC) phenomena (the oscillations of the diamagnetic moment) in a non-simply connected mesoscopic conductor. PC is produced by a sensibility of a single particle wave function to a force-free field which is taken into account via the twisted boundary conditions:

$$\Psi|_{\varphi=0} = e^{i\Delta\varphi}\, \Psi|_{\varphi=2\pi},\qquad(1)$$

where $\varphi$ is an angular variable, and $\Delta\varphi$ is the topological phase shift.

In multiparticle systems, Eq. (1) results in the oscillatory dependence of the thermodynamic and kinetic characteristics on $\Delta\varphi$. If $\Delta\varphi$ is governed by any varying external parameter $\gamma$, e.g. the magnetic field, the response of a system is the oscillatory function of $\gamma$.

Normally, the actual experiments are performed on thin quasi-one-dimensional submicron metallic or semiconductor based [1] loops pierced by a magnetic flux $\Phi$, and $\Delta\varphi$ reveals the Aharonov-Bohm effect (ABE) [2,3]:

$$\Delta\varphi_{AB} = \frac{q}{\hbar c}\oint \mathbf{A}\cdot d\mathbf{l} = 2\pi\frac{\Phi}{\Phi_0},\qquad(2)$$

where $\Phi_0 = hc/q$, q is the charge of a conduction particle[*].

The ABE-oscillations of the diamagnetic moment (the PCs) reflect the broken clockwise-anticlockwise symmetry of charge carriers momenta caused by the external vector potential. The PC is defined as

$$I_{PC} = -c(\partial F/\partial \Phi),\qquad(3)$$

where $F$ is the free energy of the loop.

---

[*] The external flux is swept adiabatically slowly with time, and, in fact, one observes oscillations with a certain temporal period connected with $\Phi_0$.

Vitaliy A. Cherkaskiy, Sergey N. Shevchenko, Alexander S. Rozhavsky, and Israel D. Vagner

In a recent publication [4], we have proposed a new mechanism for the observation of adiabatically slow oscillations of PC with time specific for 2D quantum Hall systems. This mechanism does not apply to an external magnetic field, i.e., the oscillatory current $I_{PC}$ (Eq. (3)) exists even at $\Phi = 0$ [4]. The bottomline physics behind these spontaneous PC can be understood along the following lines.

The time reversal symmetry breaking is in general achieved as the combined action of the ABE and topological spin-orbit interactions. It was shown in [5] that the topological phase shift $\Delta\varphi$ in Eq. (1) is the sum of the ABE, Aharonov-Casher [6] and Berry [7] phases providing the topologically nontrivial spatial charge carriers spins distribution:

$$\Delta\varphi = \Delta\varphi_{AB} + \Delta\varphi_{AC} + \Delta\varphi_{B} , \quad (4)$$

where

$$\Delta\varphi_{AC} = \frac{\mu_B}{\hbar c} \oint d\mathbf{l} \cdot (\mathbf{E} \times \sigma) , \quad (5)$$

$$\Delta\varphi_B = s\pi(1 - \cos\chi) , \quad (6)$$

here $\mu_B$ is the Bohr magneton, $\mathbf{E}$ is the electric field, $\sigma$ is the charge carrier spin vector, $\chi$ is the tilt angle of a magnetic texture, $s = \pm 1$ is the spin projection on a magnetic field.

As $(d/d\Phi)(\Delta\varphi) = 2\pi/\Phi_0$, the PC can be a nonzero function of $\Delta\varphi_{AC} + \Delta\varphi_B$ even at $\Delta\varphi_{AB} = 0$.

It was proposed in Ref. 4 to create a spatial distribution of the charge carriers spins through the hyperfine interaction with polarized nuclei. The contact hyperfine interaction is [8]

$$H_{hf} = \frac{8\pi}{3} g\mu_B\mu_n \sum_i \mathbf{I}_i \cdot \sigma \delta(\mathbf{r} - \mathbf{R}_i) , \quad (7)$$

where $\mu_n$ is the nuclear magneton, g is the g-factor, $\mathbf{I}_i$, $\sigma$, $\mathbf{R}_i$, $\mathbf{r}$ are the nuclear and the charge carrier spins and the position vectors, respectively. Once the nuclear spins are polarized, i.e., if $\langle\sum_i \mathbf{I}_i\rangle \neq 0$, the charge carriers feel the effective field $\mathbf{B}_{hf}(t)$ which lifts the spin degeneracy even in the absence of an external magnetic field. At low temperatures, the nuclear relaxation rates are inconveniently small [9], in particular in GaAs/AlGaAs the nuclear spin relaxation times are of the order of $10^3$ sec [10]. The Zeeman splitting reaches one tenth of the Fermi energy [11,12]. The Aharonov-Casher phase (5) arises from the spin-orbit interaction [13] which in GaAs/AlGaAs 2D-gas has the form [14]

$$H_{SO} = \frac{\alpha}{\hbar} (\sigma \times \mathbf{p}) \cdot \nu , \quad (8)$$

where $\mathbf{p}$ is the charge carrier momentum, $\nu$ is the normal to the surface, $\alpha = 0.25 \cdot 10^{-9}$ eV·cm for holes [14] and $\alpha = 0.6 \cdot 10^{-9}$ eV·cm for electrons [15], and

$$\Delta\varphi_{AC} = \frac{m^*}{\hbar^2} \oint \alpha(\nu \times \sigma) \cdot d\mathbf{l} . \quad (9)$$

The combination $\Delta\varphi_{AC} + \Delta\varphi_B$ itself does not depend on $\mathbf{B}_{hf}$ explicitly, the oscillatory dependence on $\mathbf{B}_{hf}(t)$ emerges in the PC through the mesoscopic factor $\cos(2\pi\sqrt{\mu_s(\mathbf{B}_{hf})/\Delta})$ [5] where $\mu_s(\mathbf{B}_{hf})$ is the Zeeman shifted chemical potential of the charge carriers with the spin projection s, and $\Delta$ is the spacing between the quantized electron levels in a 1D-ring. The effect of Berry phase in 2DEG based AlSb/InAs/AlSb heterostructure was observed in [16].

In this paper we consider PC in two cases: i) when nuclei are polarized along a certain direction in the plane and ii) when nuclear spins form an out-of-plane crown texture. We show that:

i). In this case $\Delta\varphi_B = 0$, and $\Delta\varphi_{AC} \neq 0$ only if the spin-orbit coupling is inhomogeneous [$\alpha = \alpha(\varphi)$] such that

$$\int_0^{2\pi} d\varphi \, \alpha(\varphi) \, e^{i\varphi} \neq 0 . \quad (10)$$

The inhomogeneity of the spin-orbit coupling plays the same role as the topologically nontrivial spin texture [4,17].

ii). In this case the PC exists at $\alpha$ = const and even at $\alpha = 0$ the PC is nonzero due to $\Delta\varphi_B$.

**Calculation of persistent currents**

The induced PC is given by the Eq. (3) at $\Phi = 0$. The standard algebra (see the Appendix) gives the following equation for the $I_{PC}$:

$$I_{PC} \cong \frac{eT}{\hbar} \sum_{l=1}^{\infty} \sum_j \frac{\sin(2\pi l n_F^{(j)})}{\sinh(lT/\tilde{T}^{(j)})}\bigg|_{\Phi=0} , \quad (11)$$

where T is the temperature, index j numerates the roots of the equation:

$$\varepsilon(n_F^{(j)}) = \mu . \quad (12)$$

Here $\varepsilon$ is the eigenvalue of the Schrödinger equation, $\mu$ is the chemical potential and

$$\tilde{T}^{(j)} = \frac{1}{2\pi^2} \left|\frac{\partial \varepsilon}{\partial n}\right|_{n=n_F^{(j)}}, \quad (13)$$

is the crossover temperature.

In what follows we solve the Schrödinger equation for the charge carriers confined to a 1D-ring with the radius $\rho$, obtain $n_F^{(j)}$ and $\tilde{T}^{(j)}$, and analyze $I_{PC}$ in various geometries.

i). The in-plane polarized nuclei, $\Delta\varphi_B = 0$.

The charge carriers Hamiltonian takes the form

$$\hat{H} = \frac{\hat{p}^2}{2m^*} + \frac{1}{2\hbar}\{\alpha\sigma\cdot\mathbf{n}, \hat{p}\}_+ - g\mu_B B_{hf}\sigma_x, \quad (14)$$

here $\mathbf{B}_{hf}$ is oriented along the x-axis in the xy-plane, $m^*$ is the effective mass, g is the g-factor, $\{...\}_+$ stands for the anticommutator,

$$\sigma\cdot\mathbf{n} = \sigma_x \cos\varphi + \sigma_y \sin\varphi,$$

$$\hat{p} = -(i\hbar/\rho)(\partial/\partial\varphi - i(\Phi/\Phi_0)), \quad \Phi_0 = hc/e.$$

Consider weak spin-orbit coupling $\alpha << \Delta\rho$, where $\Delta = \hbar^2/2m^*\rho^2$.

The spectrum linear in $\alpha$ is

$$\varepsilon_n^\pm \cong \Delta\left(n - \frac{\Phi}{\Phi_0}\right)^2 \mp g\mu_B B_{hf} \pm 2\Delta\left(n - \frac{\Phi}{\Phi_0}\right)\frac{\langle\alpha\cos\varphi\rangle}{2\Delta\rho}, \quad (15)$$

where

$$\langle\alpha\cos\varphi\rangle = \frac{1}{2\pi}\int_0^{2\pi} d\varphi\, \alpha(\varphi)\cos\varphi. \quad (16)$$

The PC takes the form

$$I_{PC} \cong \frac{4\pi eT}{\hbar\Delta\rho}\langle\alpha\cos\varphi\rangle \times$$

$$\times \sum_{l=1}^\infty l\, \frac{\sin(2\pi l\sqrt{\mu/\Delta})\sin(\pi l\, b/\sqrt{\Delta\mu})}{\sinh(\pi^2\, lT/\sqrt{\Delta\mu})}, \quad (17)$$

where $b = g\mu_B B_{hf}(t)$.

At low temperatures $T << \sqrt{\Delta\mu}/\pi^2$ the r.h.s. of Eq. (17) takes the form of the series rectangles:

$$I_{PC} \cong \frac{4e}{\hbar}\sqrt{\mu/\Delta}\frac{\langle\alpha\cos\varphi\rangle}{\rho} \times$$

$$\times \left\{\tilde{\delta}\left[2\pi\sqrt{\mu/\Delta}\left(1 - \frac{b}{2\mu}\right)\right] - \tilde{\delta}\left[2\pi\sqrt{\mu/\Delta}\left(1 + \frac{b}{2\mu}\right)\right]\right\}, \quad (18)$$

where $\tilde{\delta}(x)$ is the rectangle with the height $\sqrt{\Delta\mu}/\pi T$ and the width $\pi^2 T/\sqrt{\Delta\mu} << 1$ centered at the points $x = 2\pi k$, where k is an integer. The magnitude of $I_{PC}$ (17) is of the order of

$$I_{PC} \sim \frac{\alpha}{\Delta\rho} I_0, \quad (19)$$

where $I_0 = eV_F/2\pi\rho$ is the magnitude of a normal ABE persistent current. At high temperatures, $T >> \sqrt{\Delta\mu}/\pi^2 \equiv \tilde{T}$, the PC decreases with temperature in a standard exponential way:

$$I_{PC} \cong \frac{8\pi eT}{\hbar\Delta\rho}\langle\alpha\cos\varphi\rangle \times$$

$$\times e^{-\pi^2 T/\sqrt{\Delta\mu}}\sin(2\pi\sqrt{\mu/\Delta})\sin(\pi b/\sqrt{\Delta\mu}). \quad (20)$$

In submicron rings, the opposite limit $\alpha \gtrsim \Delta\rho$ is more favorable. In this case we can perform the perturbation scheme over the «parity» of the spin-orbit coupling. One can achieve slowly varying on the scale of $k_F^{-1}$ coordinate dependent $\alpha(\varphi)$ by means of a controlled distribution of impurities. If the «even» component $\langle\alpha\cos\varphi\rangle$ is made much larger than the «odd» one $\langle\alpha\sin\varphi\rangle$ the unperturbed Hamiltonian takes the form

$$\hat{H} = \frac{\hat{p}^2}{2m^*} + \frac{1}{2\hbar}\{\alpha\sigma_x\cos\varphi, \hat{p}\}_+ - g\mu_B B_{hf}\sigma_x, \quad (21)$$

and the Schrödinger equation can be solved exactly. The spectrum is

$$\varepsilon_n^\pm = \Delta\left(n \pm \frac{\langle\alpha\cos\varphi\rangle}{2\Delta\rho}\right)^2 \mp g\mu_B B_{hf}. \quad (22)$$

The perturbation potential is

$$\hat{V} = \frac{1}{2\hbar}\{\alpha\sigma_y\sin\varphi, \hat{p}\}_+. \quad (23)$$

One can easily see that the first correction over $\hat{V}$ to the spectrum is zero, and the second correction is negligible at $b >> (\alpha/\rho)\sqrt{\mu/\Delta}$.

The PC is

$$I_{PC} \cong \frac{4eT}{\hbar} \times$$

$$\times \sum_{l=1}^\infty \frac{\sin\left(\pi l\frac{\langle\alpha\cos\varphi\rangle}{\Delta\rho}\right)\sin(2\pi l\sqrt{\mu/\Delta})\sin(\pi l\, b/\sqrt{\Delta\mu})}{\sinh(\pi^2\, lT/\sqrt{\Delta\mu})}. \quad (24)$$

In mesoscopic devices b decreases with time exponentially $b \sim b_0 \exp(-t/t_1)$. The time scale $t_1$ is macroscopically long at low temperatures [10]. The dependence of the $I_{PC}$ on b (i.e., on t) for Eq. (24) is shown at Fig. 1.

Equation (24) differs from Eq. (17) provided $\langle \alpha \cos \varphi \rangle \sim \Delta \rho$. If the fluctuating component $\langle \alpha \cos \varphi \rangle << \Delta \rho$, the result (24) is reduced to (17).

ii). The polarized nuclear spins form a crown, $\Delta \varphi_B \neq 0$. Consider $\mathbf{B}_{hf}$ directed along the cylindrically symmetric crown (Fig. 2) tilted to the z-axis by the angle $\chi$. The electron spectrum can be obtained exactly following the paper [18]. We represent the Hamiltonian in the form

$$\hat{H} = \Delta \left( -i \frac{\partial}{\partial \varphi} + \frac{\Phi}{\Phi_0} \right)^2 + \frac{1}{2} \left\{ \begin{pmatrix} 0 & e^{-i\varphi} \\ e^{i\varphi} & 0 \end{pmatrix}, \right.$$
$$\left. \frac{\alpha}{\rho} \left( -i \frac{\partial}{\partial \varphi} + \frac{\Phi}{\Phi_0} \right) - b \sin \chi \right\}_+ - b \cos \chi \, \sigma_z .$$
(25)

The solution to the spectral equation is

$$\Psi = \begin{pmatrix} \Psi_1 \, e^{i(m-1/2)\varphi} \\ \Psi_2 \, e^{i(m+1/2)\varphi} \end{pmatrix}, \quad (26)$$

where $m = n + 1/2$, n is an integer.

The spectrum is

$$\varepsilon_m^\pm = \Delta \left( \lambda_m^2 + \frac{1}{4} \right) \pm \left[ \left( \Delta^2 + \frac{\alpha^2}{\rho^2} \right) \lambda_m^2 + 2\Delta\kappa b\lambda_m + b^2 \right]^{1/2}, \quad (27)$$

where $\lambda_m = m + (\Phi/\Phi_0)$, $\kappa = \cos \chi - (\alpha/\Delta\rho) \sin \chi$. The dependence of the PC (Eq. (36)) on b calculated numerically with the spectrum (27) is plotted at Fig. 3.

After some straightforward but cumbersome calculations we get at $\kappa < 1$ the asymptotic expression for $I_{PC}$:

$$I_{PC} \cong \frac{4eT}{\hbar} \sum_{l=1}^{\infty} (-1)^l \frac{\sin\left(\pi l \kappa / \sqrt{1 + (\mu\alpha^2/\Delta\rho^2 b^2)}\right) \sin(2\pi l \sqrt{\mu/\Delta}) \sin(\pi l \, b/\sqrt{\Delta\mu})}{\sinh(\pi^2 \, lT/\sqrt{\Delta\mu})} . \quad (28)$$

Consider the case of a «strong» Zeeman splitting $b >> (\alpha/\rho) \sqrt{\mu/\Delta}$. In this limit one can easily see that

$$(-1)^l \sin(\pi l \kappa) = \sin(\Delta\varphi_{AC} + \Delta\varphi_B), \quad (29)$$

where

$$\Delta\varphi_{AC} = 2\pi \frac{\alpha m^*}{\hbar^2} \rho \sin \chi . \quad (30)$$

Equation (30) is evidently obtained from the general definition (5) by the substitutions

$$(\mathbf{v} \times \boldsymbol{\sigma}) \to \sigma_\rho \to \sin \chi , \quad (31)$$

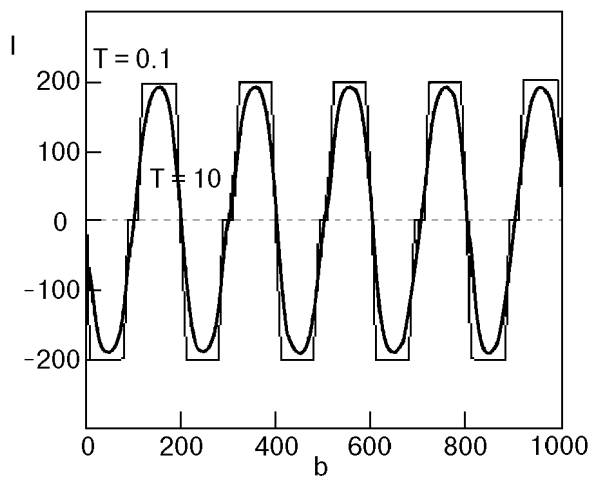

Fig. 1. Dependence of $I_{PC}$ upon b when nuclei are polarized along a certain direction in the plane for $\mu \sim 10^4$, $\alpha/\rho \sim 2$. ($\mu$, $\alpha/\rho$, b and T are expressed here in units of $\Delta$; I in units of $I_0$; for real submicron rings $\Delta = 10^{-3}$–$10^{-2}$ K.)

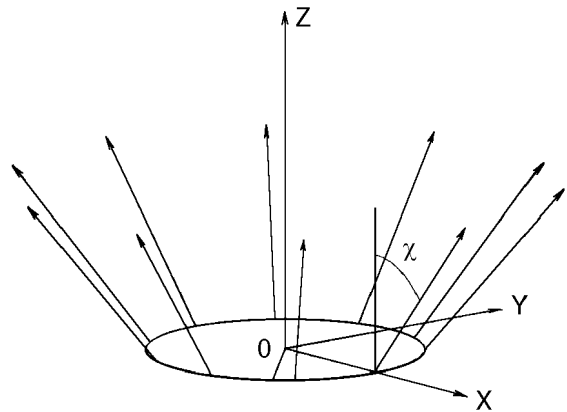

Fig. 2. $\mathbf{B}_{hf}$ directed along the cylindrically symmetric crown tilted to the z-axis by the angle $\chi$.

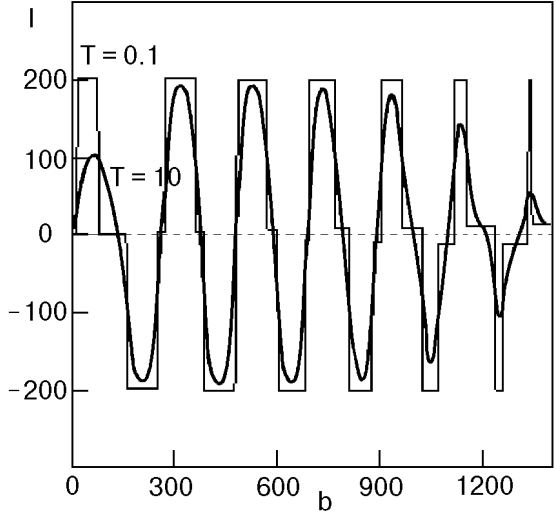

Fig. 3. Dependence of $I_{PC}$ upon b in the case when nuclear spins form an out-of-plane crown texture. ($\kappa \sim 0.5$, $\mu \sim 10^4$, $\alpha/\rho \sim 2$, quantities b, T, $\mu$, $\alpha/\rho$ are expressed here in units of $\Delta$; I in units of $I_0$.)

which means that in a «strong» hyperfine field all the spins are aligned to the crown direction.

In a «weak» hyperfine field $b \ll (\alpha/\rho)\sqrt{\mu/\Delta}$, the topological phases are expressed in terms of averaged spin components:

$$\Delta\varphi_{AC} = 2\pi \frac{\alpha m^*}{\hbar^2} \rho \langle\sigma_\rho\rangle, \qquad (32)$$

where $\langle\sigma_\rho\rangle \cong b \sqrt{\Delta/\mu}\,(\rho/\alpha) \sin\chi \ll 1$, and

$$\Delta\varphi_B = \pi(1 - \cos\chi)\langle\sigma_\chi\rangle, \qquad (33)$$

where $\langle\sigma_\chi\rangle \cong b \sqrt{\Delta/\mu}\,(\rho/\alpha) \ll 1$.

The remarkable feature of the Eq. (28) is that $I_{PC} \neq 0$ even at $\alpha = 0$ when $\Delta\varphi_B \neq 0, \pm\pi$, i.e. at $\chi \neq 0, \pi/2$, (mod $\pi$).

## Appendix

In this chapter, following the lines of [19] we perform the derivation of the PC for the arbitrary dispersion law of charge carriers.

At a fixed chemical potential, the PC is

$$I_{PC} = \sum_{n=-\infty}^{\infty} \frac{i_n}{e^{(\varepsilon_n - \mu)/T} + 1}, \qquad (A.1)$$

where $i_n$ is the partial current of the n-th orbital:

$$i_n = \frac{e}{\hbar} \frac{\partial \varepsilon}{\partial n}. \qquad (A.2)$$

Making use of the Poisson summation formula were present the r.h.s. in (A.1) in the form

$$I_{PC} = \frac{e}{\hbar} \int_{-\infty}^{\infty} dn \frac{\partial \varepsilon/\partial n}{e^{(\varepsilon(n)-\mu)/T} + 1} +$$

$$+ 2 \frac{e}{\hbar} \sum_{l=1}^{\infty} \int_{-\infty}^{\infty} dn \frac{(\partial \varepsilon/\partial n)\cos(2\pi n l)}{e^{(\varepsilon(n)-\mu)/T} + 1} =$$

$$= -\frac{eT}{\hbar} \log\left(e^{-(\varepsilon(n)-\mu)/T} + 1\right)\Big|_{-\infty}^{\infty} +$$

$$+ 4\pi T \frac{e}{\hbar} \sum_{l=1}^{\infty} \sum_{\text{Im}(n_k)>0} \text{Im}\, e^{2\pi i l n_k}, \qquad (A.3)$$

where $\varepsilon(n_k) = \mu + 2i\pi T(2k-1)$, which gives at $T \ll \mu$:

$$n_k \cong n_F + \frac{i\pi T(2k-1)}{\partial \varepsilon/\partial n|_{n=n_F}}. \qquad (A.4)$$

Eventually, we arrive to Eqs. (11)–(13):

$$I_{PC} \cong \frac{eT}{\hbar} \sum_{l=1}^{\infty} \sum_j \frac{\sin(2\pi l n_F^{(j)})}{\sinh(lT/\tilde{T}^{(j)})}\bigg|_{\Phi=0}, \qquad (A.5)$$

where

$$\tilde{T}^{(j)} = \frac{1}{2\pi^2}\left|\frac{\partial\varepsilon}{\partial n}\right|_{n=n_F^{(j)}}, \quad \varepsilon(n_F^{(j)}) = \mu. \qquad (A.6)$$

Since $\tilde{T}^{(j)}$ for various j vary insignificantly*, we safely replace $\tilde{T}^{(j)}$ by some averaged $\tilde{T} \cong \tilde{T}^{(j)}$ and get

$$I_{PC} \cong \frac{eT}{\hbar} \sum_{l=1}^{\infty} \sinh^{-1}\left(\frac{lT}{\tilde{T}}\right) \sum_j \sin(2\pi l n_F^{(j)}). \qquad (A.7)$$

---

* E.g. for in-plane $\mathbf{B}_{hf}$-configuration $\tilde{T}^{\pm} = \sqrt{\mu \pm b}$, so the difference between different $\tilde{T}^{(j)}$ is of the order of $B_{hf}/\mu$.